\documentclass[intlimits,twoside,a4paper]{article}

\usepackage[cp1251]{inputenc}
\usepackage{epsf,multicol,ifthen}

\usepackage{cmpj3}

\usepackage{bm}

\issue{2019}{22}{1}{13702}
\doinumber{10.5488/CMP.22.13702}

\title[Mechanisms of electron scattering]%
{Mechanisms of electron scattering in uniaxially deformed $\textit{n}$-$\text{Ge} \langle \text{Sb, Au}\rangle$ single crystals}
\author[S.V. Luniov \textsl{et al.}]{S.V. Luniov\refaddr{lntu}, P.F.~Nazarchuk\refaddr{lntu}, A.I.~Zimych\refaddr{lntu}, 
Yu.A.~Udovytska\refaddr{lntu}, O.V.~Burban\refaddr{nuft}}
\addresses{
\addr{lntu} Lutsk National Technical University, 75 Lvivska St., 43018 Lutsk, Ukraine  
\addr{nuft} Volyn College of the National University of Food Technologies,
6 Cathedral St., 43016 Lutsk, Ukraine 
}

\date{Received January 5, 2019, in final form February 4, 2019}

\begin{document}

\maketitle

\begin{abstract}
Temperature dependencies for concentration and the Hall mobility of electrons for the $\textit{n}$-$\text{Ge} \langle \text{Sb}\rangle$ and \linebreak  $\textit{n}$-$\text{Ge} \langle \text{Sb, Au}\rangle$ single crystals uniaxially deformed along the crystallographic directions [100] and [111] are obtained on the basis of piezo-Hall effect measurements. A deformation-induced increase of the Hall mobility of electrons for $\textit{n}$-$\text{Ge} \langle \text{Sb, Au}\rangle$ single crystals at the uniaxial pressure along the crystallographic direction [100] has been revealed. A comparison of the obtained experimental results with the corresponding theoretical calculations of temperature dependencies of the Hall mobility showed that the obtained effect occurs at the expense of the reduction probability of electron scattering on the fluctuational potential. Its amplitude depends on the tempe\-rature and on the value of the uniaxial pressure. It has also been shown that an increase of the Hall mobility for the $\textit{n}$-$\text{Ge} \langle \text{Sb, Au}\rangle$ single crystals uniaxially deformed along the crystallographic direction [111] with an increasing temperature turns out to be insignificant and is observed only for the uniaxial pressures $P<0.28$~GPa. A decrease of the Hall mobility of electrons at the expense of the deformational redistribution of electrons among the valleys of the germanium conduction band with different mobility should be taken into account in the present case. The Hall mobility magnitude for the uniaxially deformed $\textit{n}$-$\text{Ge} \langle \text{Sb}\rangle$ single crystals is determined only by the mechanisms of phonon scattering and we have not observed the effect of the growth of the Hall mobility with an increase of temperature or the magnitude of uniaxial pressure. This demonstrates a secondary role of the fluctuational potential in the present case. 
\pacs 72.20.Fr, 72.10.-d
\end{abstract}

\section{Introduction}

The development of modern electronics is associated with the search and development of new materials or the improvement of the properties of the existing ones. At present, a semiconductor material such as germanium is a promising material for creating various electronic devices and sensors such as diodes, triodes, dosimeter devices, transducers, Hall sensors, detectors of infrared radiation~\cite{Selesniov, claes}. Technologies of strained germanium find their practical use in nanosized transistor structures and nanophotonics~\cite{kobayashi, kobayashi2009, choi2007, petykiewicz2016, boztug2014}. The use of $n$-$\text{Ge}$ single crystals as a material for channels NMOSFET transistors allows one to increase their gain factor and the tunnel current in relation to those transistors whose channel is made of $n$-$\text{Si}$ \cite{kobayashi, kobayashi2009, choi2007}. The purposeful impact of Germanium lattice deformation opens prospects for the creation of fundamentally new elements of nanophotonics on its basis~\cite{petykiewicz2016, boztug2014}. However, almost all scientific publications on nanoelectronics, in particular, works \cite{kobayashi, kobayashi2009, choi2007, petykiewicz2016, boztug2014} which are devoted to the investigation and modelling of physical properties of strained germanium and nanostructures on its basis are restricted only to the examination of the deformation impact and the related effects on the atoms of the germanium lattice. However, the impact of the defective structure of germanium single crystals, created by controllable and uncontrollable impurities, was not taken into account. These impurities are introduced at the synthesis of such nanostructures and determine the degree of compensation for the present single crystals. Such impurities can create both shallow and deep energy levels in the band gap of germanium and significantly affect the functional characteristics of semiconductor devices on its basis and also largely determine the percentage of the output of suitable products. Theoretical models of the deformation influence on different kinetic and optical effects are very scarce both for germanium and in other semiconductors in the presence of deep energy levels of impurities. Today, the theory of deep levels in semiconductors, unlike the theory of shallow levels is based only on their semi-empirical models use \cite{baranskyj2}. The practical value of such investigations is connected with the fact that the impurity centers with deep energy levels determine the light-emitting diodes spectra. They are centers of rapid recombination, create additional regions of photosensitivity, strongly affect the tensosensitivity of semiconductors. Therefore, investigations of the impact of the alloyed impurities with deep levels on electrical and optical properties of the deformed germanium single crystals are urgent both from fundamental and applied points of view. Surely, these investigations, in their turn, will allow scientists and engineers to provide certain scientific and me\-tho\-di\-cal recommendations concerning technological conditions of synthesis and modelling of germanium properties and strained nanostructures on its basis.

\section{Results and discussion}

	\begin{figure}[!b]
	\centerline{
		\includegraphics[width=0.4\textwidth]{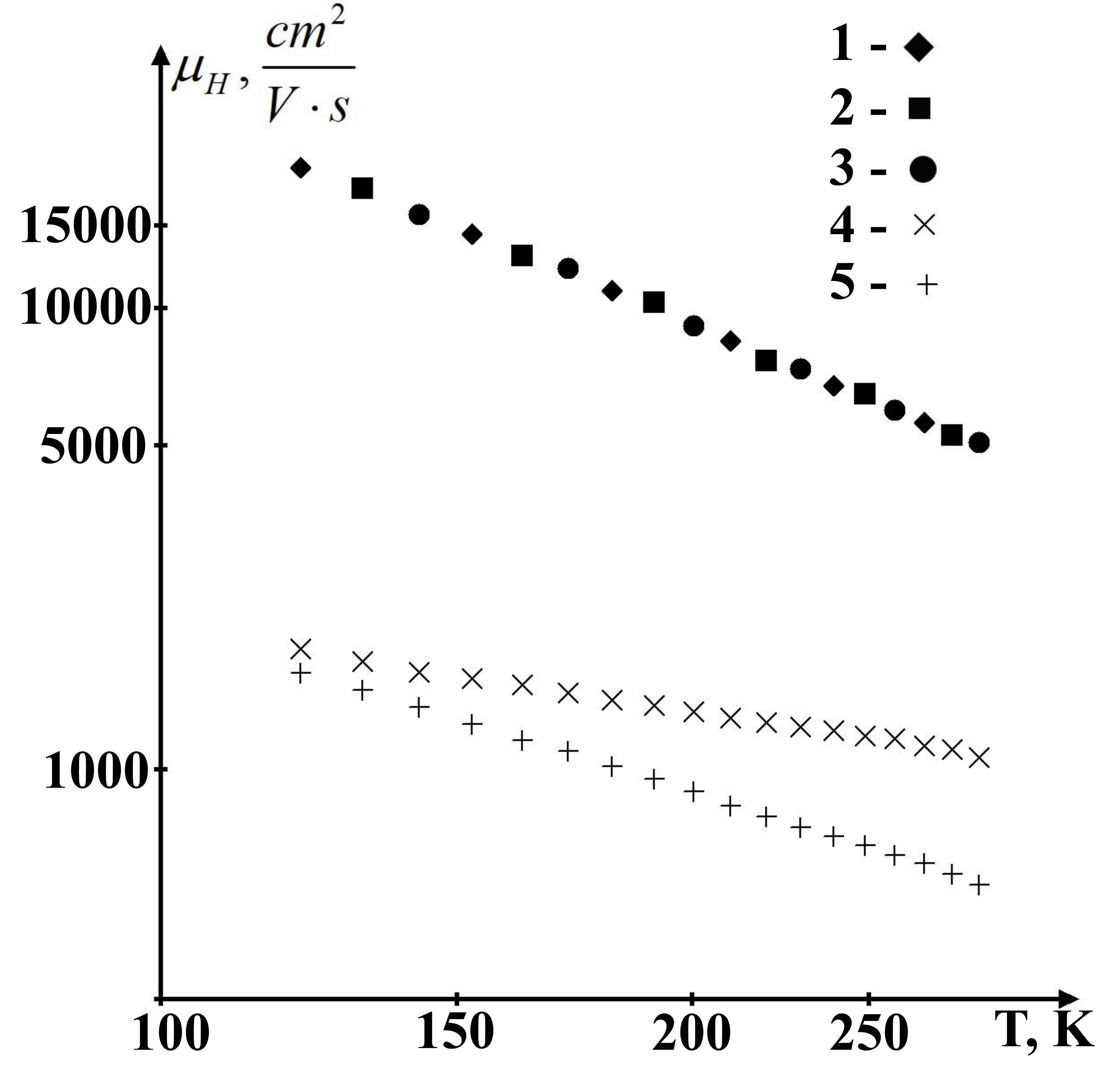}
	}
	\caption{Temperature dependencies of the Hall mobility for the $\textit{n}$-$\text{Ge} \langle \text{Sb}\rangle$ single crystals at different values of the uniaxial pressure along the crystallographic directions $[100]$ and $[111]$: 1 -- $0$~GPa; 2 -- $0.34$~GPa (for the crystallographic direction [100]); 3 -- $0.85$~GPa (for the crystallographic direction [100]); 4 -- $0.47$~GPa (for the crystallographic direction [111]); 5 -- $0.96$~GPa (for the crystallographic direction [111]).}
	\label{fig:1}
\end{figure}

\begin{figure}[!t]
		\begin{multicols}{2}
		\centerline{
		\includegraphics[width=0.4\textwidth]{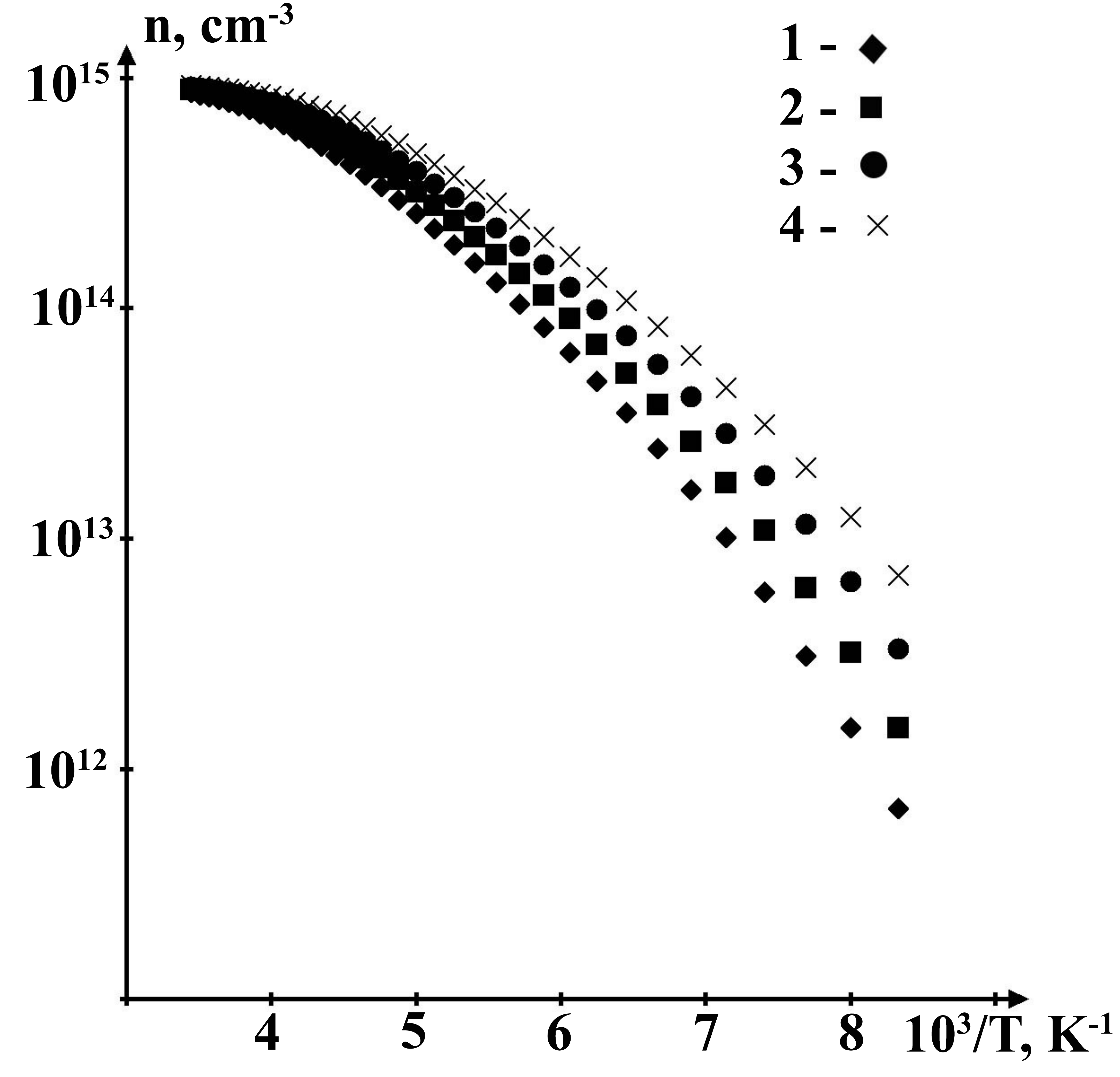}
}
		\caption{Temperature dependencies of the concentration of electrons for the $\textit{n}$-$\text{Ge} \langle \text{Sb, Au}\rangle$
		at diffe\-rent values of the uniaxial pressure along the crystallographic direction $[100]$: 1 -- $0$~GPa; 2 -- $0.29$~GPa; 3~--~$0.59$~GPa; 4 -- $0.88$~GPa.
		}
		\label{fig:2}
		        	\centerline{
			        \includegraphics[width=0.4\textwidth]{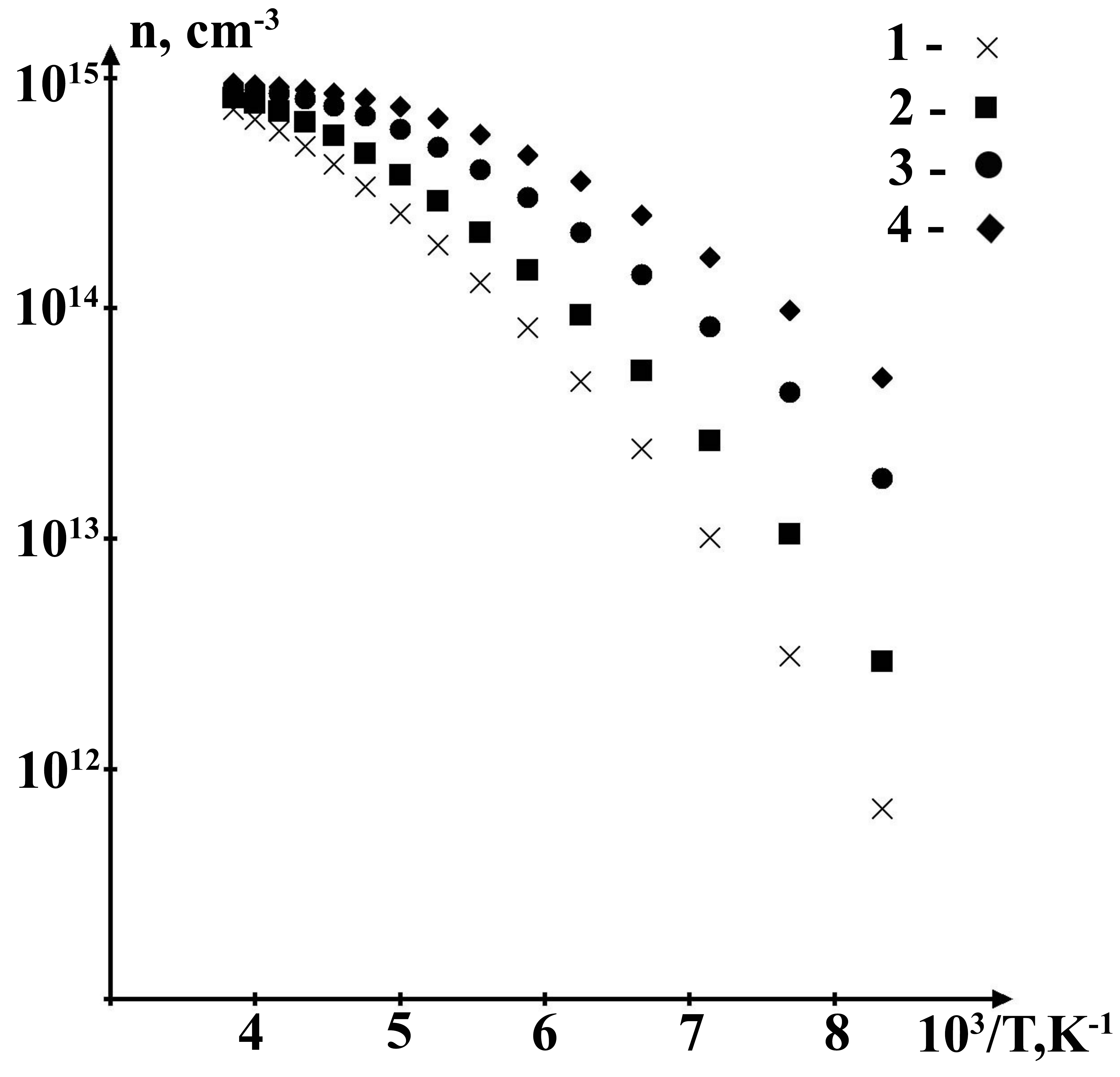}
}
			        \caption{Temperature dependencies of the concentration of electrons for the $\textit{n}$-$\text{Ge} \langle \text{Sb, Au}\rangle$ 
			        at diffe\-rent values of the uniaxial pressure along the crystallographic direction $[111]$: 1 -- $0$~GPa; 2 -- $0.28$~GPa; 3~--~$0.69$~GPa; 4 -- $0.97$~GPa.
		          	}
		         	\label{fig:3}
		\end{multicols}
	\end{figure}

So, we were measuring the temperature dependences of the Hall constant and the electrical con\-duc\-ti\-vi\-ty for uniaxially deformed $n$-$\text{Ge}$ single crystals, alloyed by Sb and Au impurities. Uniaxial deformation was attached along the crystallographic directions [100] and [111]. 
The crystallographic orientation of the investigated single crystals was determined using an X-ray machine, with an accuracy of $10'$.
Germanium samples for research were cut in the form of a rectangular parallelepipeds,  $0.8\times 0.8\times 10~\text{mm}$ in size. Measurements were conducted for $n$-$\text{Ge}$ samples of two groups. $n$-$\text{Ge} \langle \text{Sb} \rangle$ samples of the first group were alloyed only by Sb impurity, $N_\text{Sb}=5\cdot 10^{14}~\text{cm}^{-3}$ concentration. $\textit{n}$-$\text{Ge} \langle \text{Sb, Au}\rangle$ samples of the second group were alloyed by Sb impurity, $N_\text{Sb}=9.8\cdot 10^{14}~\text{cm}^{-3}$ concentration and Au impurity, $N_\text{Au}=5.05\cdot 10^{14}~\text{cm}^{-3}$ concentration. The temperature control was carried out with the help of the copper-constantan differential thermocouple. During the experiment, the accuracy of the measurement of temperature was $\pm 1$~K.   As it is known \cite{milnes1973}, impurity of Sb is a shallow donor. It forms in the band gap of germanium shallow energy level $E_C-0.0096~\text{eV}$. Au impurity forms a deep donor level $E_V+0.04~\text{eV}$ and three acceptor levels $E_V+0.15~\text{eV}$, $E_C-0.2~\text{eV}$ and $E_C-0.04~\text{eV}$ in Ge. The piezo-Hall effect study for the $n$-$\text{Ge} \langle \text{Sb} \rangle$ single crystals uniaxially deformed along the crystallographic directions $[100]$ and $[111]$ shows that concentration of electrons at temperatures $T>77$~K is equal to the concentration of alloying impurity  Sb and does not depend on the temperature. This is explained by the fact that the shallow donors of Sb will be completely ionized for such temperatures. Experimental results of temperature dependencies of the Hall mobility for $\textit{n}$-$\text{Ge} \langle \text{Sb}\rangle$ single crystals uniaxially deformed along the crystallographic directions $[100]$ and $[111]$ are presented in figure~\ref{fig:1}. As can be seen in figure~\ref{fig:1}, the Hall mobility at the deformation along the crystallographic direction $[100]$ does not depend on the magnitude of the uniaxial pressure. There is no deforming redistribution of electrons between the minima of germanium conduction band which will be shifting upward (according to the scale of energies) at the deformation with the same speed for the present case \cite{herring}. The deforming redistribution of electrons between three minima of germanium conduction band with higher mobility (which ascend upwards) and one minimum with smaller mobility (which descends down according to the scale of energies at the deformation) will take place at the uniaxial deformation $n$-$\text{Ge} \langle \text{Sb} \rangle$ along the crystallographic directions $[111]$. In this case, the concentration of electrons in three minima with higher mobility will decrease. However, the concentration of electrons in the minima with smaller mobility will increase \cite{herring}. This will lead to a decrease of the average Hall mobility of electrons with an increase of the uniaxial pressure magnitude. This fact explains the obtained experimental results (figure~\ref{fig:1}, curves 4 and 5). The growth of the concentration of electrons in the germanium conduction band with an increasing temperature for $\textit{n}$-$\text{Ge} \langle \text{Sb, Au}\rangle$ single crystals (figure~\ref{fig:2} and figure~\ref{fig:3}) is explained by thermal ionization of the deep acceptor level of gold $E_C-0.2~\text{eV}$. 

	
	Moreover, for the given single crystals, as can be seen in figure~\ref{fig:2} and figure~\ref{fig:3} (curves 2$-$4), the concentration of electrons increases when the magnitude of the uniaxial pressure increases. This is associated with a decrease of ionization energy of the level $E_C-0.2~\text{eV}$ at the deformation \cite{semenyuk}.
	
	 A research of the hydrostatic pressure influence on the position for the deep level of gold $E_C-0.2~\text{eV}$ in germanium single crystals has been conducted by the authors of works \cite{holland1962, daunov2001}. The baric coefficient magnitude for ionization energy of the given level which had been calculated in work \cite{holland1962}, turned out to be understated, taking into account the experimental results of work \cite{daunov2001}. This is explained by the fact that the authors of work \cite{holland1962} did not take into account the impact of fluctuation potential in their calculations \cite{daunov2001, kamilov2008}. As it is known \cite{shklovsky, karpov, bezludniy, luniov}, such a potential arises in the alloyed compensated semiconductors or in semiconductors with radiation-induced defects when the concentration of free charge carriers is small in comparison with the concentration of ionized impurity centres or charged radiation defects. The amplitude of this potential can be quite significant which will lead to modulation of density of states for charge carriers and, as a result, to a decrease of Hall mobility. For the given case, Hall mobility of electrons can be written as follows  \cite{bezludniy, luniov}:
\begin{equation} \label{eq1}
\mu_\text{H}=\mu_{\text H_0}\exp\left( -\frac{\Delta}{kT}\right),
\end{equation}
where $\mu_{\text H_0}$  is Hall mobility for uncompensated semiconductor,  $\Delta$ is amplitude of fluctuation potential. In accordance with \cite{karpov},
\begin{equation} \label{eq2}
\Delta=\frac{q^2N^{\frac{2}{3}}}{\varepsilon n^{\frac{1}{3}}}\,,
\end{equation}
where $N$ is the total concentration of charged impurities or defects,  $\varepsilon$ is dielectric penetration,  $n$ is the concentration of electrons in the conduction band, $q$  is electron charge.

\begin{figure}[!t]
			\begin{multicols}{2}
			\centerline{
			\includegraphics[width=0.4\textwidth]{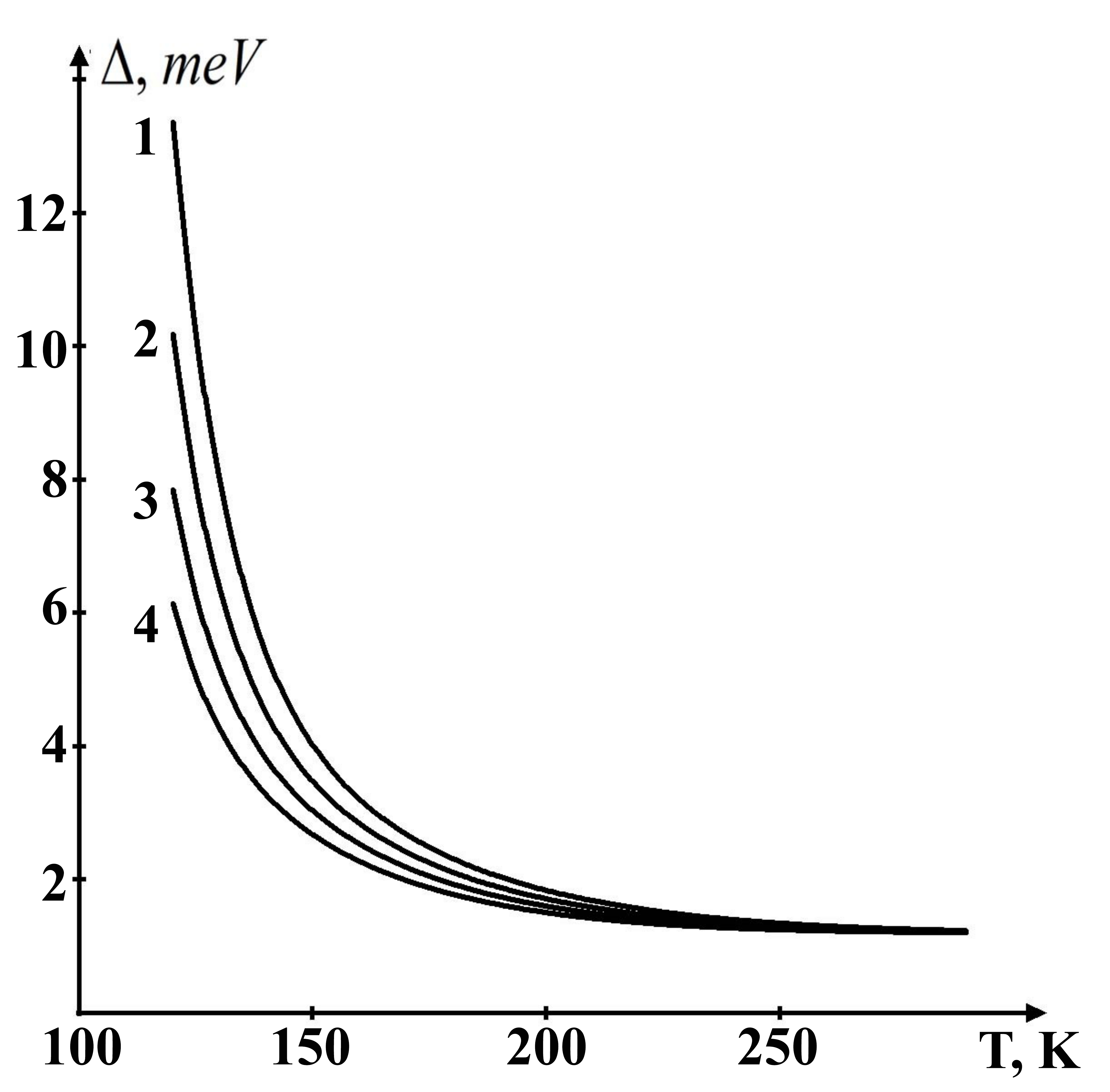}
}
			\caption{Temperature dependencies of the amplitude of fluctuation potential for the uniaxially deformed $\textit{n}$-$\text{Ge} \langle \text{Sb, Au}\rangle$ single crystals at different values of the uniaxial pressure along the crystallographic direction $[100]$: 1 -- $0$~GPa; 2 -- $0.29$~GPa; 3~-- $0.59$~GPa; 4 -- $0.88$~GPa.
			}
			\label{fig:4}

			\centerline{
			\includegraphics[width=0.4\textwidth]{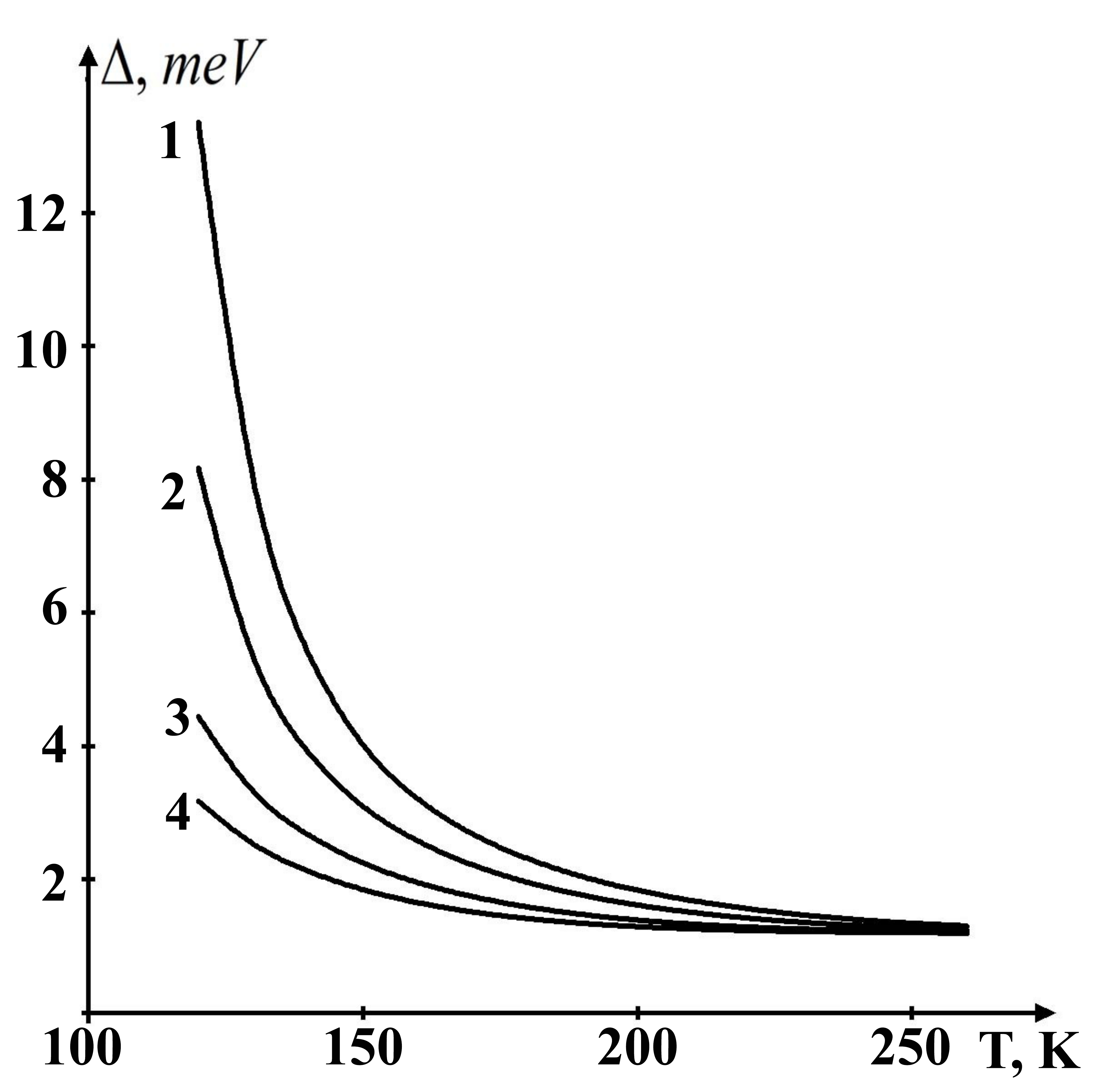}
}
			\caption{Temperature dependencies of the amplitude of fluctuation potential for the uniaxially deformed $\textit{n}$-$\text{Ge} \langle \text{Sb, Au}\rangle$ single crystals at different values of the uniaxial pressure along the crystallographic direction $[111]$: 1 -- $0$~GPa; 2 -- $0.28$~GPa; 3~-- $0.69$~GPa; 4 -- $0.97$~GPa.
			}
			\label{fig:5}
		\end{multicols}
		\end{figure}

The reduction of the amplitude of fluctuation potential for the uniaxially deformed $\textit{n}$-$\text{Ge} \langle \text{Sb, Au}\rangle$ single crystals (figure~\ref{fig:4} and figure~\ref{fig:5}), taking into account the expression (\ref{eq2}) and experimental results (figure~\ref{fig:2} and figure~\ref{fig:3}), is caused by an increase of the concentration of electrons in the germanium conduction band at the expense of magnification of the temperature or magnitude of the uniaxial pressure. 

\begin{figure}[!t]
			\begin{multicols}{2}
			\centerline{
			\includegraphics[width=0.41\textwidth]{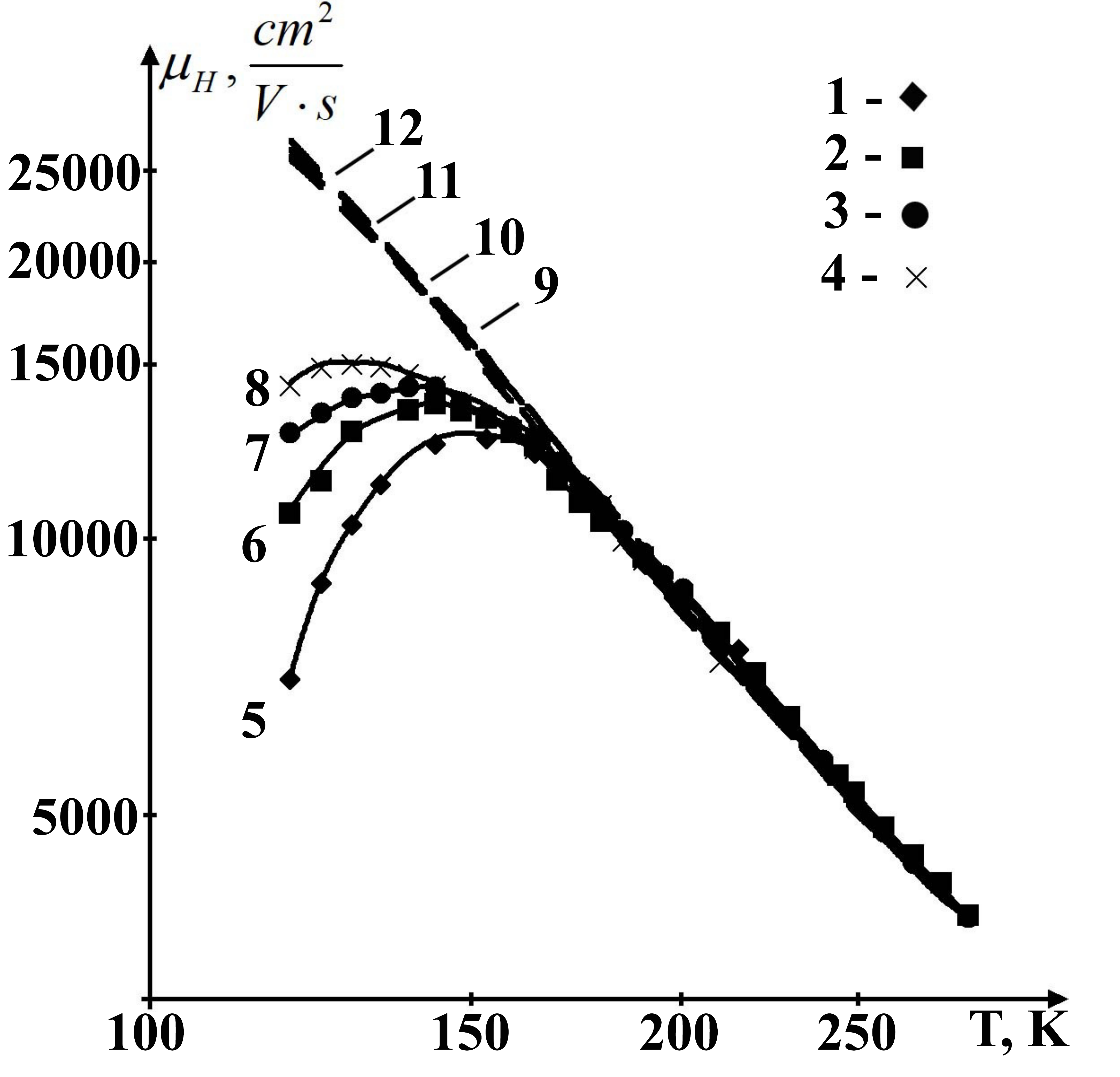}
}
			\caption{Temperature dependencies of the Hall mobility for the $\textit{n}$-$\text{Ge} \langle \text{Sb, Au}\rangle$ single crystals at different values of the uniaxial pressure along the crystallographic directions $[100]$: 1 -- $0$~GPa, 2 -- $0.29$~GPa, 3 -- $0.59$~GPa, 4 -- $0.88$~GPa (experimental results); 5 -- $0$~GPa, 6 -- $0.29$~GPa, 7 -- $0.59$~GPa, 8~-- $0.88$~GPa (solid curves are theoretical calculations taking into account the fluctuation potential); 9 -- $0$~GPa, 10~-- $0.29$~GPa, 11 -- $0.59$~GPa, 12 -- $0.88~$GPa (dashed curves are theoretical calculations without taking into account the fluctuation potential).
			}
			\label{fig:6}
		\centerline{
						\includegraphics[width=0.4\textwidth]{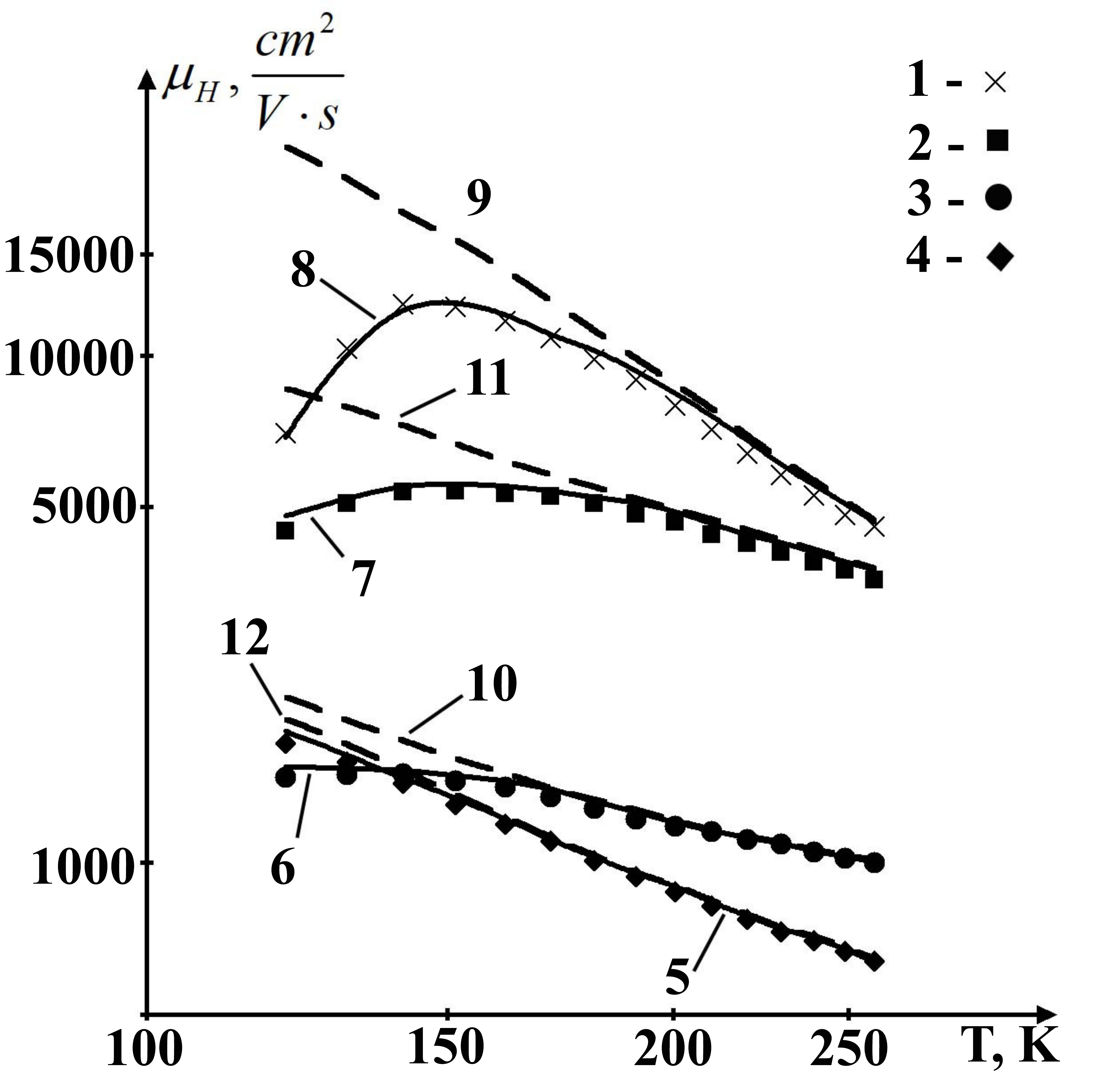}
		}
	    	\caption{Temperature dependencies of the Hall mobility for the $\textit{n}$-$\text{Ge} \langle \text{Sb, Au}\rangle$ single crystals at differ\-ent values of uniaxial pressure along the crystal\-lographic directions $[111]$: 1 -- $0$~GPa, 2 -- $0.28$~GPa, 3 -- $0.69$~GPa, 4 -- $0.97$~GPa (experimental results); 5 -- $0.97$~GPa, 6 -- $0.69$~GPa, 7 -- $0.28$~GPa, 8~-- $0$~GPa (solid curves are theoretical calculations taking into account fluctuation potential); 9 -- $0$~GPa, 10 -- $0.69$~ GPa, 11 -- $0.28$~GPa, 12 -- $0.97$~GPa (dashed curves are theoretical calculations without taking into account the fluctuation potential).
		}
		\label{fig:7}
	\end{multicols}
	\end{figure}

	This, in its turn, explains the anomalous temperature dependencies of Hall mobility for the \linebreak $\textit{n}$-$\text{Ge} \langle \text{Sb, Au}\rangle$ single crystals uniaxially deformed along the crystallographic direction $[100]$ (figure~\ref{fig:6}).
	
	 As it follows from figure~\ref{fig:6}, Hall mobility does not depend on uniaxial pressure at high temperatures because the amplitude of fluctuation potential is small and weakly depends on deformation (figure~\ref{fig:4}). For the case of uniaxial pressure along the crystallographic direction $[111]$ (figure~\ref{fig:7}), the deformation-induced increase of the Hall mobility at low temperatures was not observed. Only at $T<140$~K and $P>0.69$~GPa (figure~\ref{fig:7}, curves 3 and 4) Hall mobility slightly increased when uniaxial pressure grew. In this case, as mentioned above, it is also necessary to take into account a decrease of the Hall mobility of electrons at the expense of the deformation redistribution of electrons between the valleys of the germanium conduction band with different mobility. Corresponding theoretical calculations which are presented in figure~\ref{fig:6} and figure~\ref{fig:7} (solid and dashed curves) were carried out by us for the quantitative estimation of the Hall mobility magnitude for the uniaxially deformed $\textit{n}$-$\text{Ge} \langle \text{Sb, Au}\rangle$ single crystals.


According to \cite{bir}, the isoenergetic surfaces of the germanium conduction band are ellipsoids of rotation, with an axis of symmetry which coincides with the crystallographic direction $[111]$. Then, the mobility of charge carriers for an arbitrary direction can be determined from the ratio: 
\begin{equation} \label{eq3}
\mu=\mu_\bot\sin^2\theta+\mu_\parallel\cos^2\theta\,,
\end{equation}
where  $\theta$ is the angle between the examined direction and the main axis of the ellipsoid; $\mu_\bot$  and $\mu_\parallel$ is the mobility of charge carriers across and along the axis of the ellipsoid.

In accordance with (\ref{eq1}), for $n$-$\text{Ge}$ single crystals undeformed and those uniaxially deformed along the crystallographic direction $[100]$ 
\begin{equation} \label{eq4}
\mu_{0}=\frac{1}{3}\mu_\parallel+\frac{2}{3}\mu_\bot\,.
\end{equation}

Under uniaxial pressure $n$-$\text{Ge}$ along the crystallographic direction $[111]$, one minimum i.e., the main axis of the isoenergetic ellipsoid which is oriented along the axis of deformation, will descend according to the scale of energies by a value \cite{bir}
\begin{equation} \label{eq5}
\Delta E_1=-\left(\Xi_\text{d}+\frac{1}{3}\Xi_\text{u} \right)\left(S_{11}+2S_{12} \right)P-\frac{1}{3}\Xi_\text{u}S_{44}P  ,
\end{equation}
and the other three minima will ascend by a value
\begin{equation} \label{eq6}
\Delta E_2=-\left(\Xi_\text{d}+\frac{1}{3}\Xi_\text{u} \right)\left(S_{11}+2S_{12} \right)P+\frac{1}{9}\Xi_\text{u} S_{44}P  ,
\end{equation}
which will lead to the emergence of the energy gap between them  $\Delta E_{1,2}=\frac{4}{9}\Xi_\text{u} S_{44}P$.

If $n_1$ is a concentration of electrons in a descending minimum, and  $n_2$ in three minima which ascend under uniaxial pressure $P\parallel J\parallel [111]$, then the total concentration of electrons in the conduction band of germanium is as follows:
\begin{equation} \label{eq7}
n=n_1+n_2\,.
\end{equation}

For the nondegenerate electron gas \cite{kireev}
\begin{equation} \label{eq8}
n_1=2\left( \frac{2\piup m_1kT}{\hbar^2}\right)^{\frac{3}{2}}\text{e}^{\frac{E_\text{F}-\Delta E_1}{kT}},\,\,\,\,\,
n_2=2\left( \frac{2\piup m_2kT}{\hbar^2}\right)^{\frac{3}{2}}\text{e}^{\frac{E_\text{F}-\Delta E_2}{kT}} .
\end{equation}

Then,
\begin{equation} \label{eq9}
\frac{n_2}{n_1}=\left(\frac{m_2}{m_1} \right)^{\frac{3}{2}}\text{e}^{-\frac{\Delta E_{1,2}}{kT}}=A\,,
\end{equation}
where $m_1$, $m_2$  represent the effective mass of the density of states for the given minima.
For the isoenergetic surface, which is an ellipsoid of rotation, the effective mass of the density of states is as follows:
\begin{equation} \label{eq10}
m=N^{\frac{2}{3}}\left(m_\parallel m^2_\bot \right)^{\frac{1}{3}}   ,
\end{equation}
where $N$  denotes the number of equivalent ellipsoids,  $\mu_\parallel$ and $m_\bot$ are components for the tensor of the effective mass of an electron along and across the axis of the ellipsoid. 

In accordance with the (\ref{eq3}), the mobility of electrons under the uniaxial pressure $P\parallel J\parallel [111]$ in the descending minimum is equal to
\begin{equation} \label{eq11}
\mu_1=\mu_\parallel\,.
\end{equation}

Hence, for the three minima which ascend according to the scale of energies,
\begin{equation} \label{eq12}
\mu_2=\frac{8\mu_\bot+\mu_\parallel}{9}\,.
\end{equation}

From expressions (\ref{eq7}) and (\ref{eq9}) we find
\begin{equation} \label{eq13}
n_1=\frac{n}{A+1}\,,\,\,\,\,\,n_2=\frac{An}{A+1}\,.
\end{equation}

Then, for an arbitrary value of the uniaxial pressure $P\parallel J\parallel [111]$, the conductivity $n$-$\text{Ge}$ is as follows:
\begin{equation} \label{eq14}
\sigma_P=qn\mu=q\left(n_1\mu_1+n_2\mu_2 \right).
\end{equation}

Taking into account expressions (\ref{eq11})$-$(\ref{eq14}), the mobility of electrons at the uniaxial pressure $n$-$\text{Ge}$ along the crystallographic direction $[111]$ is equal to
\begin{equation} \label{eq15}
\mu=\frac{\mu_1+A\mu_2}{A+1}\,.
\end{equation}

As it is known \cite{luniov2014}, electron scattering on optical phonons is also possible for germanium, in addition to their scattering on acoustic phonons and ions of impurity. Electron scattering on optical phonons is caused by the interaction of electrons with phonons, whose frequencies correspond to temperature $T_\text{C1}=430$~K (intravalley scattering) and intervalley scattering on acoustic phonons with characteristic temperaturies  $T_\text{C2}=320$~K. 

Intervalley electron scattering and electron scattering on optical phonons are described by a scalar relaxation time $\tau_j$:

\begin{equation} \label{eq16}
\frac{1}{\tau_j}=a_j\varphi_j\,,
\end{equation}
where
\begin{align*}
 a_j&=\displaystyle \frac{\Xi_j^2\big(m_\text{d}^j\big)^{\frac{3}{2}}}{\sqrt{2}\piup\rho \hbar^2\big(kT_{\text{C}j}\big)^{\frac{1}{2}}  }\left(\frac{T}{T_{\text{C}j}}\right)^{\frac{1}{2}}, \nonumber \\ 
\varphi_j(x)&=\displaystyle \frac{1}{\re^\frac{T_{\text{C}j}}{T}-1}\left[\left(x+\frac{T_{\text{C}j}}{T} \right)^{\frac{1}{2}}+\re^{\frac{T_{\text{C}j}}{T}} \theta \left(x,\frac{T_{\text{C}j}}{T} \right) \left( x-\frac{T_{\text{C}j}}{T} \right)^{\frac{1}{2}} \right], 
\end{align*} 
 \noindent $m_\text{d}^j$ is the effective mass of the density of states for electrons of the conduction band, $\Xi_j$  is a constant of intervalley or optical deformation potential; $\rho$  is a density of the crystal; $T_{\text{C}j}$  is characteristic temperature of $j$ phonon;  $x=\frac{\varepsilon}{kT}$ is a dimensionless energy of electron; $\theta\big( x;\frac{T_{\text{C}j}}{T}\big) $  is a step function. 
For intervalley scattering 
\begin{equation} \label{eq17}
m_\text{d}^j=\big(m_{\parallel j}m^2_{\bot j}\big)^{\frac{1}{3}}\big(Z_j-1\big)\,,
\end{equation}
where $m_{\parallel j}$, $m_{\bot j}$  are a longitudinal and transverse component of the tensor of the effective mass for electrons which are in the ellipsoid  of $j$  type; $Z_j$  is the number of equivalent ellipsoids of the conduction band of  $j$ type.

For intravalley electron scattering on optical phonons
\begin{equation} \label{eq18}
m_\text{d}^j=(m_{\parallel j}m^2_{\bot j})^{\frac{1}{3}}Z_j^{\frac{2}{3}},
\end{equation}

Expressions for components of the relaxation-time tensor $\tau_\parallel^{a,j}$  and  $\tau_\bot^{a,j}$ under conditions of mixed electron scattering on acoustic phonons and ions of impurity \cite{baranskyj} can be written based on the theory of anisotropic scattering:
\begin{equation} \label{eq19}
\tau_\parallel^{a,i}=\frac{a_\parallel}{\sqrt{k_\text{B}}T^{\frac{3}{2}}}\cdot \frac{x^{\frac{3}{2}}}{x^2+b_0}\,,\,\,\,\,\,
\tau_\bot^{a,i}=\frac{a_\bot}{\sqrt{k_\text{B}} T^{\frac{3}{2}}}\cdot\frac{x^{\frac{3}{2}}}{x^2+b_1}\,.
\end{equation}

(Expressions for $a_\parallel$ , $a_\bot$ , $b_0$ , $b_1$ are presented in appendix \ref{appendix_A}).

Then, in the most general case, the scattering of electrons on the acoustic phonons, ions of impurity, optical phonons and acoustic phonons, which are responsible for the intervalley electron scattering, expressions for components of the relaxation-time tensor can be presented as follows \cite{luniov2014}:

\begin{equation} \label{eq20}
\frac{1}{\tau_\parallel}=\frac{1}{\tau_\parallel^{a,i}}+\frac{1}{\tau_1}+\frac{1}{\tau_2}\,,\,\,\,\,\,\,\frac{1}{\tau_\bot}=\frac{1}{\tau_\bot^{a,i}}+\frac{1}{\tau_1}+\frac{1}{\tau_2}\,,
\end{equation}

 \noindent where $\tau_\parallel^{a,j}$ , $\tau_\bot^{a,i}$  , $\tau_1$  , $\tau_2$  are longitudinal and transverse components of the relaxation-time tensor for scattering on acoustic phonons and ions of impurity, respectively;  $\tau_1$  , $\tau_2$  are the relaxation time for intervalley scattering and scattering on optical phonons.

	Components of tensors of mobility can be expressed through components of tensors of relaxation times and effective mass:
\begin{equation} \label{eq21}
\mu_\parallel=\frac{e}{m_\parallel}\left\langle \tau_\parallel\right\rangle,\,\,\,\,\, 
\mu_\bot=\frac{e}{m_\bot} \left\langle \tau_\bot\right\rangle,
\end{equation}
\begin{equation} \label{eq22}
\langle\tau_\parallel\rangle=\frac{4}{3\sqrt{\piup}}\int\limits_{0}^{\infty}\rd xx^{\frac{3}{2}}\re^{-x}\tau_\parallel\,,\,\,\,\,\,  
\langle\tau_\bot\rangle=\frac{4}{3\sqrt{\piup}}\int\limits_{0}^{\infty}\rd xx^{\frac{3}{2}}\re^{-x}\tau_\bot\,.
\end{equation}

Numerical values of parameters for the energy-band structure of germanium single crystals such as components of tensors of the acoustic potential of deformation and effective mass ($\Xi_\text d=-6.4~\text{eV}$,  $\Xi_\text{u}=16.4~\text{eV}$,  $m_\parallel=1.58m_0$, $m_\bot=0.082m_0$) \cite{baranskyj}, constants of electron-phonon interaction for optical  $\Xi_{430}=4\cdot 10^8~\frac{\text{eV}}{\text{cm}}$  and  acoustic phonons $\Xi_{320}=1.4\cdot 10^8~\frac{\text{eV}}{\text{cm}}$,  which are responsible for the intervalley electron scattering \cite{luniov2014}, should be taken into account for carrying out theoretical calculations. Temperature dependencies of the Hall mobility for the uniaxially deformed $\textit{n}$-$\text{Ge} \langle \text{Sb, Au}\rangle$ single crystals (which were obtained based on the given calculations) are presented in figure~\ref{fig:6} and figure~\ref{fig:7} (solid and dashed curves). 

The results of comparison of the curves, which were obtained with and without taking into account fluctuation potential (figure~\ref{fig:6}, curves 5$-$12), show that under low temperatures the Hall mobility significantly depends on the magnitude of the amplitude of the given potential. For the $\textit{n}$-$\text{Ge} \langle \text{Sb, Au}\rangle$ single crystals uniaxially deformed along the crystallographic direction $[111]$ the effect of the growth of the Hall mobility with an increasing temperature is not significant and is observed only for the uniaxial pressures $P<0.28$~GPa (figure~\ref{fig:7}, curves 1 and 2). In the given case, the magnitude of the Hall mobility is determined by the deformational reconstruction of the germanium conduction band and the effectiveness of electron scattering on the fluctuation potential. The change of relative contribution of the given mechanisms under an increasing uniaxial pressure explains the insignificant growth of the Hall mobility under $P>0.69$~ GPa and temperatures $T<140$~K (figure~\ref{fig:7}, curves 3 and 4).

\section{Conclusions}
	Electrons scattering on the ions of shallow Sb impurities for the germanium single crystals studied is described using the Coulomb screening potential of the impurity. For deep impurities Au, this approach does not apply, since today there are no adequate theoretical models of deep centres in semiconductors. Therefore, it is difficult to make any estimates of the influence of such impurities on the mechanisms of electron scattering in $\textit{n}$-$\text{Ge} \langle \text{Sb, Au}\rangle$ single crystals. However, for the investigated impurity concentrations Sb and Au, the electron scattering on the ions of the gold impurity can be considered secondary, since the experimental dependences of the Hall mobility of the electrons on the temperature for the uniaxially deformed $\textit{n}$-$\text{Ge} \langle \text{Sb, Au}\rangle$ single crystals are well described based on the proposed theoretical model of mobility. Additional doping of $n$-$\text{Ge} \langle \text{Sb} \rangle$ single crystals by the impurity of gold leads to an increase in the degree of compensation of such single crystals and consequently to the reduction of the screening effect. The reduction of the effect of screening is caused by the occurrence of large-scale fluctuations of the concentration of charged ions of doping impurities and, accordingly, the fluctuation potential, whose amplitude depends on temperature and uniaxial pressure. Therefore, description of various electron transfer phenomena in $\textit{n}$-$\text{Ge} \langle \text{Sb, Au}\rangle$ single crystals, for the investigated temperatures and concentrations of doping impurities Sb and Au, can be restricted to the mechanisms of electron scattering on the acoustic and optical phonons (intravalley scattering), acoustic phonons, which are responsible for intervalley scattering, ions of shallow Sb impurities and fluctuation potential.

	The obtained experimental results and theoretical calculations show that for single crystals \linebreak $\textit{n}$-$\text{Ge} \langle \text{Sb, Au}\rangle$ uniaxially deformed along the crystallographic direction [100], a decrease of the amplitude of fluctuation potential under an increase of temperature or the magnitude of uniaxial pressure causes a growth of the Hall mobility. Its further decrease, passing through the maximum with an increasing temperature, is explained by the increase of probability of electron scattering on optical phonons and phonons, which are responsible for the intervalley electron scattering. Scattering of electrons on the fluctuation potential is secondary in this case. The present mechanism of scattering does not appear for the whole area of the investigated temperatures, and the Hall mobility magnitude is fully determined by the mechanisms of the phonon scattering. The impact of deformational reconstruction of the germanium conduction band (in addition to a change of the amplitude of the fluctuation potential at the uniaxial pressure) should be additionally taken into account for the case of the single crystals $\textit{n}$-$\text{Ge} \langle \text{Sb, Au}\rangle$ uniaxially deformed along the crystallographic direction [111]. This mechanism causes a decrease of the Hall mobility. The obtained temperature dependencies of the Hall mobility are determined by different relative contributions of the given mechanisms depending on the magnitude of the uniaxial pressure.

	It is important to consider the impact of fluctuation potential on the mechanisms of electron transport in germanium both at the absence and in presence of deformation fields in modelling and developing on its basis the electronic devices and sensors, nanostructures, alloyed by different impurities (alloyed quantum dots Ge, heterostructures SiGe).

\appendix
\section{Necessary  appendices to calculate the relaxation 
	time}
\label{appendix_A}

$$
a_\parallel=\frac{\piup C_{11}\hbar^4}{k\Xi^2_\text{d} \sqrt{2m_\parallel m^2_\bot}}\cdot\frac{1}{\Phi_{0a}}\,,\,\,\,\,\, a_\bot=\frac{\piup C_{11}\hbar^4}{k\Xi^2_\text{d} \sqrt{2m_\parallel m^2_\bot}}\cdot\frac{1}{\Phi_{1a}}\,,
$$

$$
b_0=\frac{a_\parallel\cdot\Phi_{0i}}{\sqrt{k}T^{\frac{3}{2}}\tau_{0i}\left( kT\right) }\,,\,\,\,\,\, 
b_1=\frac{a_\bot\cdot\Phi_{1i}}{\sqrt{k}T^{\frac{3}{2}}\tau_{0i}\left( kT\right) }\,,
$$

$$
\tau_{0i}(kT)=\frac{\sqrt{2}m_\bot\varepsilon^2\left( kT\right) ^{\frac{3}{2}}}{\piup N_\text{d} e^4 \sqrt{m_\parallel}}\,,
$$

$$
\Phi_{1a}=1+\frac{1+\beta^2}{\beta^2}\bigg[ 2+\frac{3}{\beta^2}-\frac{3\big( 1+\beta^2\big) }{\beta^3}\alpha\bigg] \frac{\Xi_\text{u}}{\Xi_\text{d}}+\frac{\left( 1+\beta^2\right) }{\beta^4}\frac{\Xi_\text{u}^2}{\Xi_\text{d}^2}\left(A+B \right),
$$

$$A= \big( 1+\beta^2\big) \bigg[ 1+\frac{15}{4\beta^2}-\frac{3}{4\beta^3}\big( 5+3\beta^2\big) \alpha\bigg]\,, $$
$$B=\frac{C_{11}}{4C_{44}}\bigg[ -13-\frac{15}{\beta^2}+\frac{3\left( 1+\beta^2\right) }{\beta^3}\big( 5+\beta^2\big) \alpha\bigg]\,,  $$

$$
\Phi_{0a}=1+\frac{2\left( 1+\beta^2\right) }{\beta^2}\left( 1-\frac{3}{\beta^2}+\frac{3}{\beta^3}\alpha\right) \frac{\Xi_\text{u}}{\Xi_\text{d}}+\frac{\left( 1+\beta^2\right) }{\beta^4}\frac{\Xi_\text{u}^2}{\Xi_\text{d}^2}\left(D+K \right),
$$
$$D= \big( 1+\beta^2\big) \bigg[ 1-\frac{6}{\beta^2}-\frac{3}{2\beta^2\left( 1+\beta^2\right) }+\frac{15\alpha}{2\beta^3}\bigg]\,, $$
$$K=\frac{C_{11}}{C_{44}}\bigg[ 2+\frac{15}{2\beta^2}-\frac{3}{2\beta^3}\big( 5+3\beta^2\big) \alpha\bigg]\,, $$

$$
\Phi_{0i}=\frac{3}{2\beta^3}\bigg[ \left( \frac{\beta}{1+\beta^2}-\alpha\right) \ln\gamma^2-\alpha\ln\big( 1+\beta^2\big) +2L\left( a\right) +\frac{\beta\gamma^2}{2}M \bigg]\,,
$$

$$M= \frac{\beta^2-1}{\beta^2+1}+\frac{\alpha\left( \beta^2+1\right) }{\beta}\,, $$

$$
\Phi_{1i}=\frac{3}{4\beta^3} \bigg\{ \Big[ \big( 1-\beta^2\big) \alpha-\beta\Big] \ln\gamma^2+2\big( \beta^2-1\big) L\left( a\right)\bigg\}+\frac{3}{4\beta^3}R\,,
$$

$$R=-2\beta^2\alpha-\big( \beta^2-1\big) \alpha\ln\big( 1+\beta^2\big) +\frac{\gamma^2}{2}\Big[ \beta\big( 1+3\beta^2\big) +\alpha\big( 3\beta^4+2\beta^2-1 \big)\Big]\,,$$

$$
\alpha= \text{arctg}\beta\,,\,\,\,\,\, \beta^2=\frac{m_\parallel-m_\bot}{m_\bot}\,,\,\,\,\,\, \gamma=\sqrt{\frac{\piup\hbar^2 e^2 n}{2m_\parallel\varepsilon kT}}\,.
$$
$
L(a)=-\int\nolimits_{0}^{a}\ln \cos\varphi \rd\varphi
$ is the Lobachevsky function, 
$N_\text{d}$ is concentration of donor impurity,
$n$ is concentration of electrons in the conduction band.



\begin{thebibliography}{10}

\bibitem{Selesniov} Selezenev~A.A., Aleinikov~A.Yu., Ermakov~P.V., Ganchuk~N.S., Ganchuk~S.N., Jones~R.E., Phys. Solid State, 2012, \textbf{54}, No.~3, 462--467, \doi{10.1134/S1063783412030286}.
\bibitem{claes} Claeys C., Simoen E., Germanium-Based Technologies: From Materials to Devices, Elsevier Science, 2007.
\bibitem{kobayashi} Kobayashi M., Irisawa T., Magyari-Kope B., Saraswat K., Wong H.-S.P., Nishi Y., IEEE Trans. Electron Devices, 2010, \textbf{57}, No.~5, 1037--1046, \doi{10.1109/TED.2010.2042767}.
\bibitem{kobayashi2009} Kobayashi M., Irisawa T., Magyari-Kope B., Sun Y., Saraswat K., Wong H.-S.P., Pianetta P., Nishi Y., In: Proceedings of 2009 Symposium on VLSI Technology, Honolulu, HI, USA, 2009, 76--77.   
\bibitem{choi2007} Choi Y.S., Lim J.-S., Numata T., Nishida T., Thompson S.E., J. Appl. Phys., 2007, \textbf{102}, No.~10, 104507, \\ \doi {10.1063/1.2809374}.
\bibitem{petykiewicz2016} Petykiewicz J., Nam D., Sukhdeo D.S., Gupta S., Buckley S., Piggott A.Y., Vu\v{c}kovi\'{c} J., Saraswat K.C., Nano Lett., 2016, \textbf{16}, No.~4, 2168--2173, \doi {10.1021/acs.nanolett.5b03976}.
\bibitem{boztug2014} Boztug C., S\'{a}nchez-P\'{e}rez J.R., Cavallo F., Lagally M.G., Paiella R., ACS Nano, 2014, \textbf{8},  No.~4, 3136--3151, \\ \doi {10.1021/nn404739b}.
\bibitem{baranskyj2} Baranskii P.I., Fedosov A.V., Gaydar G.P., Physical Properties of Silicon and Germanium Crystals in Fields of Effective External Action, Nadstirya, Lutsk, 2000, (in Ukrainian).
\bibitem{milnes1973} Milnes A.G., Deep Impurities in Semiconductors, Wiley, New York, 1973.
\bibitem{herring} Herring C., Vogt E., Phys. Rev., 1956, \textbf{101}, 944--961, \doi{10.1103/PhysRev.101.944}.
\bibitem{semenyuk} Semenyuk A.K., Nazarchuk P.F., Fiz. Tekh. Poluprovodn., 1984, \textbf{18}, 540--542 (in Russian).
\bibitem{holland1962} Holland M.G., Paul W., Phys. Rev., 1962, \textbf{128}, No.~1, 43, \doi{10.1103/PhysRev.128.43}.
\bibitem{daunov2001} Daunov M.I., Kamilov I.K., Gabibov S.F., Semiconductors, 2001, \textbf{35}, No.~1, 59--66, \doi{10.1134/1.1340290}.
\bibitem{kamilov2008} Kamilov I.K., Daunov M.I., Gabibbov S.F., Magomedov A.B., J. Phys. Conf. Ser., 2008, \textbf{121}, 022006, \\ \doi{10.1088/1742-6596/121/2/022006}.
\bibitem{shklovsky} Shklovskii B.I., Efros A.L., Electronic Properties of Doped Semiconductors, Nauka, Moscow, 1979, (in Russian).
\bibitem{karpov} Karpov V.G., Fiz. Tekh. Poluprovodn., 1981, \textbf{15}, No.~2, 217--223 (in Russian).
\bibitem{bezludniy} Bezludnyi S.V., Kolesnikov N.V., Fiz. Tekh. Poluprovodn., 1981, \textbf{15}, 218--229 (in Russian).
\bibitem{luniov} Luniov S.V., Zimych A.I., Nazarchuk P.F., Maslyuk V.T., Megela I.G., J. Phys. Stud., 2015, \textbf{19}, No.~4, 4704.
\bibitem{bir} Bir G., Symmetry and Deformation Effects in Semiconductors, Nauka, Moscow, 1972, (in Russian).
\bibitem{kireev} Kireev P.S., Semiconductor Physics, Vysshaya Shkola, Moscow, 1969.
\bibitem{luniov2014} Luniov S.V., Burban O.V., J. Nano Electron. Phys., 2014, \textbf{6}, 01020.
\bibitem{baranskyj} Baranskii P., Buda I., Dahovskii I., Kolomoets V., Electrical and Galvanomagnetic Phenomena in Anisotropic Semiconductors, Naukova Dumka, Kiev, 1977, (in Russian).

\end{thebibliography}

%
%

\ukrainianpart

\title{Механізми розсіяння електронів в одновісно деформованих монокристалах $\textit{n}$-$\text{Ge} \langle \text{Sb, Au}\rangle$ }
\author{С.В. Луньов\refaddr{lntu}, П.Ф. Назарчук\refaddr{lntu}, А.І. Зімич\refaddr{lntu}, Ю.А. Удовицька\refaddr{lntu}, О.В. Бурбан\refaddr{nuft}}
\addresses{
\addr{lntu}Луцький національний технічний університет, вул. Львівська, 75, 43018 Луцьк, Україна
\addr{nuft} Волинський коледж Національного університету харчових технологій, вул. Кафедральна, 6, 43016 Луцьк, Україна
}
%
%
%

\makeukrtitle

\begin{abstract}
\tolerance=3000%
На основі вимірювань п'єзо-холл-ефекту одержано температурні залежності концентрації та холівської рухливості електронів для одновісно деформованих вздовж кристалографічних напрямків [100] та [111] монокристалів $\textit{n}$-$\text{Ge} \langle \text{Sb} \rangle$ та $\textit{n}$-$\text{Ge} \langle \text{Sb, Au}\rangle$. Виявлено деформаційно-індуковане зростання холівської рухливості електронів для монокристалів $\textit{n}$-$\text{Ge} \langle \text{Sb, Au}\rangle$ при одновісному тискові вздовж кристалографічного напрямку [100]. Порівняння одержаних експериментальних результатів з відповідними теоретичними розрахунками температурних залежностей холівської рухливості показали, що даний ефект виникає за рахунок зменшення ймовірності розсіяння електронів на флуктуаційному потенціалі, амплітуда якого залежить від температури та величини одновісного тиску. Показано, що зростання холівської рухливості для одновісно деформованих монокристалів $\textit{n}$-$\text{Ge} \langle \text{Sb, Au}\rangle$ вздовж кристалографічного напрямку [111] є незначним і спостерігається лише для одновісних тисків $P<0.28$~ГПа. В даному випадку необхідно також враховувати зменшення холівської рухливості електронів за рахунок деформаційного перерозподілу електронів між долинами зони провідності германію з різною рухливістю. Для одновісно деформованих монокристалів $n$-$\text{Ge} \langle \text{Sb} \rangle$ величина холівської рухливості при таких самих умовах визначається лише механізмами фононного розсіяння і ефект зростання холівської рухливості при збільшенні температури або величини одновісного тиску не спостерігався. Це свідчить про другорядну роль флуктуаційного потенціалу в даному випадку.
\keywords ефект Холла, явища переносу, електричні властивості, домішки

\end{abstract}

\end{document}